\documentclass[linenumbers,twocolumn]{aastex631}
\usepackage{amsmath}
\usepackage{graphicx}
\usepackage{rotating}
\usepackage{CJK}
 \usepackage{hyperref}

\hypersetup{
    colorlinks,
    citecolor=black,
    filecolor=black,
    linkcolor=blue,
    urlcolor=blue
}

\nolinenumbers
\renewenvironment{acknowledgments}{
    \section*{Acknowledgments}
    \nolinenumbers
}{}

\begin{document}
\begin{CJK*}{UTF8}{gbsn}
\title{Can the central compact object in HESS J1731--347 be indeed the lightest neutron star observed?}

% % Author 1
\author[0000-0003-0368-384X]{S. R. Zhang (张书瑞)}

\affiliation{Dipartimento di Fisica e Scienze della Terra, Universit\`a di Ferrara, Via Saragat 1, I--44122 Ferrara, Italy}
\affiliation{ICRANet-Ferrara, Dip. di Fisica e Scienze della Terra, Università degli Studi di Ferrara, Via Saragat 1, I-44122 Ferrara, Italy}
\affiliation{School of Astronomy and Space Science, University of Science and Technology of China, Hefei 230026, China}
\affiliation{CAS Key Laboratory for Research in Galaxies and Cosmology, Department of Astronomy, University of Science and Technology of China, Hefei 230026, China}

\correspondingauthor{S. R. Zhang}
% \correspondingauthor{J. A. Rueda}
\email{zhangsr@mail.ustc.edu.cn}

% % Author 2
\author[0000-0003-4904-0014]{J. A. Rueda}
\affiliation{ICRANet, Piazza della Repubblica 10, I-65122 Pescara, Italy}
\affiliation{ICRANet-Ferrara, Dip. di Fisica e Scienze della Terra, Universit\`a di Ferrara, Via Saragat 1, I--44122 Ferrara, Italy}
\affiliation{ICRA, Dipartamento di Fisica, Sapienza Universit\`a  di Roma, Piazzale Aldo Moro 5, I-00185 Rome, Italy}
\affiliation{Dipartimento di Fisica e Scienze della Terra, Universit\`a di Ferrara, Via Saragat 1, I--44122 Ferrara, Italy}
\affiliation{INAF, Istituto di Astrofisica e Planetologia Spaziali, Via Fosso del Cavaliere 100, 00133 Rome, Italy}
\email{jorge.rueda@icra.it}

% % Author 2
\author[0000-0002-4975-5805]{R. Negreiros}
\affiliation{Catholic Institute of Technology \\ 1 Broadway - 14th floor, Cambridge, MA 02142}
\affiliation{Instituto de F\'isica,
  Universidade Federal Fluminense, Niter\'oi, Rio de Janeiro,
  Brazil}
% \correspondingauthor{J. A. Rueda}
\email{rnegreiros@catholic.tech}

\begin{abstract}
\nolinenumbers
The exceptionally low mass of $0.77_{-0.17}^{+0.2} M_{\odot}$ for the central compact object (CCO) XMMU J173203.3–344518 (XMMU J1732) in the supernova remnant (SNR) HESS J1731–347 challenges standard neutron star (NS) formation models. The nearby post-AGB star IRAS 17287–3443 ($\approx 0.6 M_\odot$), also within the SNR, enriches the scenario. To address this puzzle, we advance the possibility that the gravitational collapse of a rotating pre-SN iron core ($\approx 1.2 M_\odot$) could result in a low-mass NS. We show that angular momentum conservation during the collapse of an iron core rotating at $\approx 45\%$ of the Keplerian limit results in a mass loss of $\approx 0.3 M_\odot$, producing a stable newborn NS of $\approx 0.9 M_\odot$. Considering the possible spin-down, this indicates that the NS is now slowly rotating, thus fulfilling the observed mass-radius relation. Additionally, the NS's surface temperature ($\approx 2 \times 10^6$ K) aligns with canonical thermal evolution for its $\approx 4.5$ kyr age. We propose the pre-SN star, likely an ultra-stripped core of $\approx 4.2 M_\odot$, formed a tidally locked binary with IRAS 17287–3443, having a 1.43-day orbital period. The supernova led to a $\approx 3 M_\odot$ mass loss, imparting a kick velocity $\lesssim 670$ km s$^{-1}$, which disrupted the binary. This scenario explains the observed 0.3 pc offset between XMMU J1732 and IRAS 17287–3443 and supports the possibility of CCOs forming in binaries, with rotation playing a key role in core-collapse, and the CCO XMMU J1732 being the lightest NS ever observed.
\end{abstract}

\keywords{Close binaries --- Stripped supernovae --- Core collapse --- Neutron stars}

%%%%%%%%%%%%%%%%%%%%%%%%%%%%%%%%%%%%%%%%%%%%%%%%%%%%%%%%%%%
%%%%%%%%%%%%%%%%%%%%%%%%%%%%%%%%%%%%%%%%%%%%%%%%%%%%%%%%%%%
\section{Introduction} \label{sec:1}
%%%%%%%%%%%%%%%%%%%%%%%%%%%%%%%%%%%%%%%%%%%%%%%%%%%%%%%%%%%
%%%%%%%%%%%%%%%%%%%%%%%%%%%%%%%%%%%%%%%%%%%%%%%%%%%%%%%%%%%

Central compact objects (CCOs) are isolated compact stars, thought to be neutron stars (NSs), located at the center of supernova remnants (SNRs) of young ages of a few thousand years \citep{2016MNRAS.458.2565D,2021MNRAS.506.5015H}, show low magnetic fields ranging $10^{10}$--$10^{11}$ G \citep{2010ApJ...709..436H,2013ApJ...765...58G,2021MNRAS.506.5015H}, and emit thermal X-ray radiation with no counterparts at any other wavelength and most do not exhibit pulsations \citep{2017JPhCS.932a2006D,2023ApJ...944...36A}. These observational properties make the study of CCOs crucial for understanding the equation of state (EOS), thermal evolution, and formation channel of NSs.

Typically, no observations suggest a binary origin for CCOs, except the CCO in the SNR HESS J1731--347 (see below), XMMU J173203.3--344518, hereafter XMMU J1732 for short. But this is not the only feature that makes XMMU J1732 an exceptional CCO study case: its mass has been recently measured to be $0.77_{-0.17}^{+0.2} M_{\odot}$ and its radius $10.4_{-0.78}^{+0.86}$ km \citep{2022NatAs...6.1444D}. The above numbers challenge the existing evolutionary channels leading to NSs, e.g., via core-collapse, given that only NS masses $\gtrsim 1.17 M_\odot$ are expected \citep{2018MNRAS.481.3305S}. This is generally consistent with observations, except for XMMU J1732. Thus, even if the general relativistic equilibrium configuration sequences of NSs allow for stable low-mass NSs, and the mass and radius of this CCO could provide new constraints on the nuclear EOS, astrophysical formation channels would avoid their formation in nature. Given the above properties of this system, it has been suggested XMMU J1732 could be a strange star \citep[see, e.g.,][]{2022arXiv221107485D, 2023A&A...672L..11H}, a hybrid star or a dark matter-admixed NS \citep[see, e.g.][]{2023ApJ...958...49S,2024PhRvD.109f3017L}.  

In this article, we argue that XMMU J1732 might indeed be a low-mass NS and advance the possible formation channel in a core-collapse scenario by considering the rotational effects of the progenitor star in a binary, which have not been previously accounted for. The presence of a (likely) post-asymptotic giant branch (post-AGB) star (IRAS 17287--3443) inside the SNR HESS J1731--347 suggests these two stars could have formed a binary that was disrupted by the SN event \citep{2016MNRAS.458.2565D}. Binaries are prevalent in the universe, and binary stellar evolution, their associated SN explosions, and products can differ significantly from those of single stars \citep{2004ApJ...612.1044P,2020A&A...637A...6L,2021A&A...656A..58L}. The gravitational and rotational effects may lead to mass transfer and rotational synchronization of the stellar components with the orbital period in compact-orbit binaries. Therefore, the pre-SN star could be a rotating core whose outermost layers may have been stripped, with the composition and structure different from the single-star case \citep{2020A&A...637A...6L,2021A&A...656A..58L}.

Bearing the above in mind, in the following sections, we explore the possibility that the pre-SN system was a tidally locked binary that led to an ultra-stripped SN, forming at its center an NS that becomes low mass owing to mass shedding, and the disruption of the binary. In section \ref{sec:2}, the observational properties of the source are introduced, a model of rotating core collapse to form an NS is constructed, and the rotational evolution of the NS leading to shedding mass during its formation is discussed. Subsequently, in section \ref{sec:3}, we argue a possible evolutionary scenario for the pre-SN binary and discuss its disruption by the SN event. Finally, we summarize and discuss our conclusions in section \ref{sec:4}.

%%%%%%%%%%%%%%%%%%%%%%%%%%%%%%%%%%%%%%%%%%%%%%%%%%%%%%%%%%
%%%%%%%%%%%%%%%%%%%%%%%%%%%%%%%%%%%%%%%%%%%%%%%%%%%%%%%%%%
\section{The low-mass NS formation and evolution} \label{sec:2}
%%%%%%%%%%%%%%%%%%%%%%%%%%%%%%%%%%%%%%%%%%%%%%%%%%%%%%%%%%
%%%%%%%%%%%%%%%%%%%%%%%%%%%%%%%%%%%%%%%%%%%%%%%%%%%%%%%%%%

%%%%%%%%%%%%%%%%%%%%%%%%%%%%%%%%%%%%%%%%%%%%%%%%%%%%%%%%%%
\subsection{Observational Properties} \label{sec:2.1}
%%%%%%%%%%%%%%%%%%%%%%%%%%%%%%%%%%%%%%%%%%%%%%%%%%%%%%%%%%

The observations of HESS J1731--347 (also known as G353.6--0.7) indicate the presence of XMMU J1732 at its center with a mass of $0.77_{-0.17}^{+0.2} M_{\odot}$ and a radius of $10.4_{-0.78}^{+0.86}$ km \citep{2022NatAs...6.1444D}. An optical star, IRAS 17287--3443, likely a post-AGB star with a mass of $0.605 M_\odot$, is also inside the SNR \citep{2016MNRAS.458.2565D}. The mass of the thick dust shell within which the optical star and the CCO are embedded has been estimated to be $1.5$--$3M_\odot$ \citep{2016MNRAS.458.2565D}. Gaia parallax measurements constrain the distance to the optical companion to be $2.5(3)$ kpc \citep{2021AJ....161..147B}. Thus, the observed angular distance of $\approx 25$ arcsecond between XMMU J1732 and IRAS 17287--3443 implies their projected separation to be $\approx 0.3$ pc. 
There is another source, HESS J1729--345, where a γ-ray excess has also been observed. This source is located near HESS J1731--347 and shares a common HII cloud association with it, but it lies approximately 1 degree away from the CCO. As explained in \cite{2018MNRAS.474..662M}, the TeV emission outside the SNR is produced by escaping cosmic rays, and HESS J1729--345 may represent a new component of target material mass.
% This distance is lower than the average separation of AGB stars in the Galaxy \citep{2002MNRAS.337..749J}, so the probability that the two objects are chance neighbors is relatively low. Therefore, they were likely members of the same binary system before the supernova explosion of the CCO’s progenitor star. JORGE: This comment doesn't seem to add much to the discussion, or I couldn't decipher it. If there is something relevant to extract from it, then uncomment it again but clarify its consequence.

%%%%%%%%%%%%%%%%%%%%%%%%%%%%%%%%%%%%%%%%%%%%%%%%%%%%%%%%%%%
\subsection{The rotating collapse of the iron core} \label{sec:2.2}
%%%%%%%%%%%%%%%%%%%%%%%%%%%%%%%%%%%%%%%%%%%%%%%%%%%%%%%%%%%

We assume XMMU J1732 is an NS formed following the traditional paradigm of gravitational collapse of the iron core of an evolved star. For simplicity, a polytropic sphere model is employed to calculate (solving the Lane-Emden equation) the density profile $\rho(R)$ of the iron core. A polytropic sphere with polytropic index $n=3$ ($\gamma=4/3$) is a suitable approximation near the critical mass, with a moment of inertia given by $I = k M R^2$, where $k=0.075$. Despite the polytropic sphere closely matching the detailed iron core model with $k=0.074$ \citep{2013ApJ...762..117B}, the spherical symmetry does not account for deformations due to rotation. This is safe for our purpose since the collapsing part is the central region of the iron core, where the oblateness is very small, making it accurate to first order. Hence, equal equatorial and polar radii ($R_{\rm eq}$ and $R_{\rm p}$) are assumed, hereafter denoted as $R$.
% \begin{figure}
% \centering\includegraphics[width=0.5\textwidth]{./figures/diagram.png}
% \caption{The diagram of the core before collapse with angular velocity $\Omega(x)$. The collapse occurs for parts with cylindrical polar coordinates less than a certain value $x$, and shedding occurs for parts greater than $x$.} 
% \label{fig:diagram}
% \end{figure}

The angular momentum of every fluid element is assumed to be conserved during the collapse. The collapsing matter is contained within a cylindrical polar coordinate in the initial model \citep[see, e.g.,][]{1985ApJ...298..474M}. For a spherical object, the mass and angular momentum contained within the cylindrical polar coordinate $x$ are
\begin{equation}
	M_b(x)=\int_{0}^{x} \int_{0}^{\sqrt{R^2-x^{'2}}} \rho\left(\sqrt{x^{'2}+z^2}\right) 4 \pi x' \, dz dx',
\label{eq:mx}
\end{equation}

\begin{equation}
	L(x)=\int_{0}^{x} \int_{0}^{\sqrt{R^2-x^{'2}}} \rho\left(\sqrt{x^{'2}+z^2}\right) 4 \pi x^{'3}\ \Omega(x') \, dz dx'.
\label{eq:Lx}
\end{equation}

When a rapidly rotating NS is formed with baryonic mass $M_b(x)$, the corresponding gravitational mass is $M(x)$. Following the general relativistic uniformly rotating NS results of \citet{2015PhRvD..92b3007C}, the relationship between the above two is $M/M_\odot\approx M_b/M_\odot-(1/20)(M_b/M_\odot)^2$, and the upper limit of angular momentum is given by 
\begin{equation}
	L_{\rm max}\approx 0.7 \frac{G M^2(x)}{c}.
	\label{eq:Lmax}
\end{equation}
Thus, the angular momentum of the original components required for the collapse into an NS must fulfill 
\begin{equation}
    L(x) \le L_{\rm max}.
    \label{eq:ineq}
\end{equation}
This necessary condition allows us to determine the critical cylindrical polar coordinate $x_{\rm crit}$, below which collapse occurs and above it mass is shed. The equality in expression (\ref{eq:ineq}) indicates the final collapsed object is at the Keplerian sequence. 

\begin{figure}
\centering\includegraphics[width=0.5\textwidth]{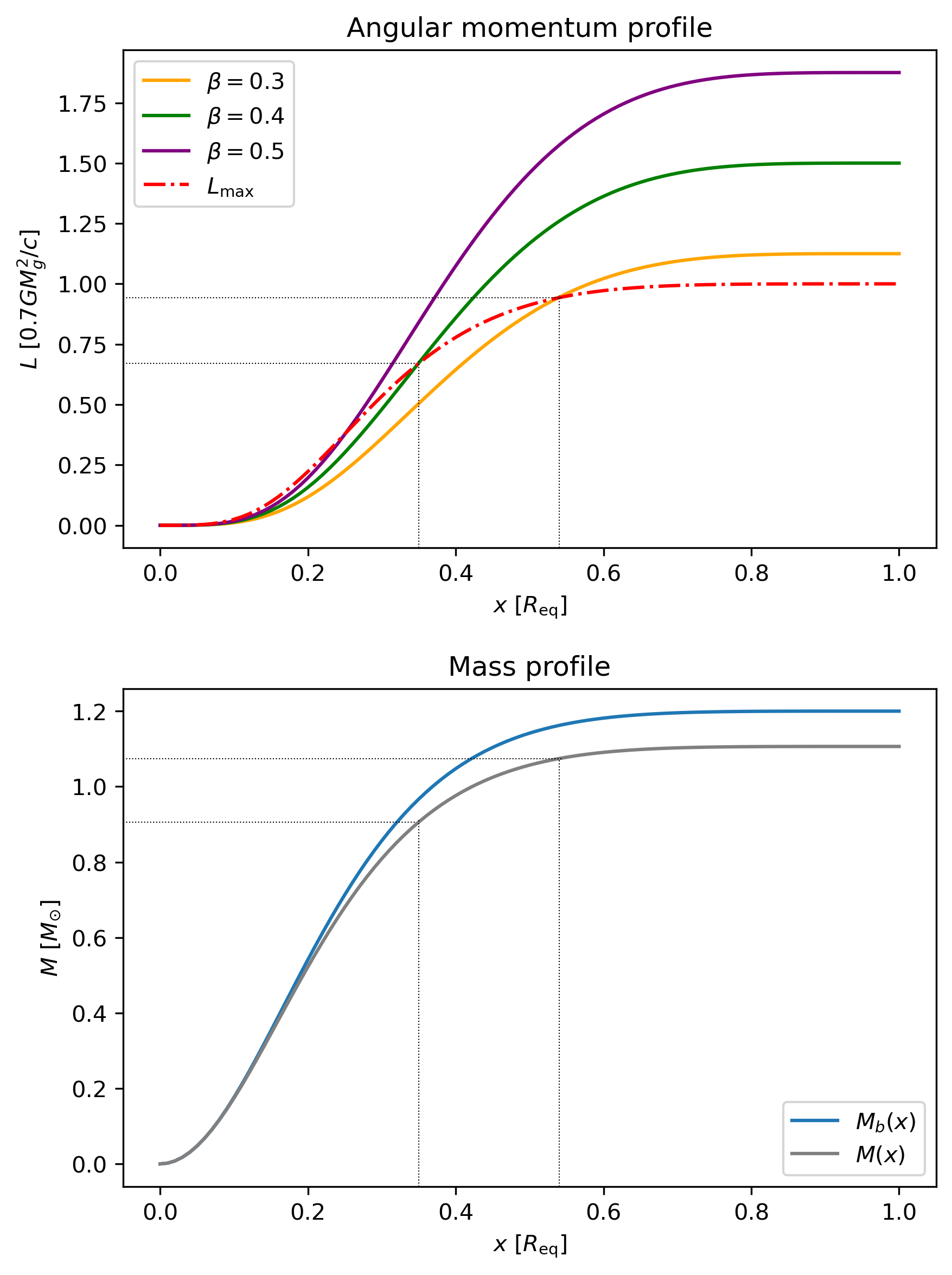}
\caption{Upper panel: angular momentum contained within the cylindrical polar coordinate $x$ (solid curves), along with the maximum angular momentum for the uniformly rotating sphere (red dash-dotted curve), for selected values of the initial angular velocity parameter $\beta$. The angular momentum is normalized to $L_{\rm max} = 0.7 G M^2/c$. Lower panel: baryonic (blue curve) and gravitational (black curve) mass contained within the cylindrical polar coordinate $x$.} 
\label{fig:unif3}
\end{figure}

We turn to exemplify the model with specific cases. Let us assume the initial iron core is uniformly rotating with an angular velocity $\Omega$ given by a fraction $\beta<1$ of the maximum angular velocity set by the mass-shedding, Keplerian limit, $\Omega_K$. According to the simulation of a uniformly rotating iron core in general relativity \citep[see, e.g.,][]{2013ApJ...762..117B}, $\Omega_K \approx 0.76 \sqrt{G M/R^3}$, so $\Omega = \beta\,\Omega_K$. We adopt an initial mass $M=1.2M_\odot$ and radius $R=2686$ km, consistent with the structure parameters of the maximum mass configuration of an iron core in the same general relativistic treatment. For rotation parameters $\beta=0.3$, $\beta=0.4$, and $\beta=0.5$, the corresponding angular momenta contained within the cylindrical polar coordinate $x$ are depicted in Fig. \ref{fig:unif3}, along with the maximum allowable angular momentum. Additionally, the lower panel of Fig. \ref{fig:unif3} gives the masses contained within the cylindrical polar coordinate $x$.

In the case of $\beta=0.3$ and $\beta=0.4$, the critical coordinate is $x_{\text{crit}}=0.54$ and $x_{\text{crit}}=0.35$ respectively, and the corresponding mass that can collapse to form an NS is $M(x_{\text{crit}}) \approx 1.07M_\odot$ and $M(x_{\text{crit}}) \approx 0.91M_\odot$. The case of $\beta=0.4$ is suitable for XMMU J1732. The actual rotation parameter $\beta$ may be slightly larger since the central core may not be a pure iron core, so its mass may be somewhat larger than $1.2M_\odot$, and the mass of the newly formed NS may be even smaller initially, increasing to its current mass through the accretion of shedding material. 
However, the exact amount of shed material accreted onto the NS and the unbound portion depends on factors like magnetic field strength, rotation, and the NS collapse process \citep[e.g.][]{2024arXiv241010938C}. While dynamic evolution of shed materials merit further numerical simulation, they are beyond the scope of this study.
The case $\beta=0.5$ leads to $L(x)$ above $L_{\rm max}$ for the appropriate NS mass values. Therefore, values $\beta \sim 0.4$--$0.45$ are reasonable.

To summarize, the collapse of an iron core of $M_b\approx M_g =1.2M_\odot$, uniformly rotating at $\Omega = 0.4 \Omega_K \approx 0.87$ rad s$^{-1}$, so a rotation period $P\approx 7.2$ s, could lead to an NS of baryonic mass $M_b\approx 0.96 M_\odot$, so a gravitational mass $M\approx 0.91 M_\odot$. In the collapse process, $\Delta M = 0.24 M_\odot$ are shed, in addition to the ejected outermost layers of the pre-SN star hosting the iron core.

%%%%%%%%%%%%%%%%%%%%%%%%%%%%%%%%%%%%%%%%%%%%%%%%%%%%%%%%%%
\subsection{Spin-down of the newborn NS} \label{sec:2.3}
%%%%%%%%%%%%%%%%%%%%%%%%%%%%%%%%%%%%%%%%%%%%%%%%%%%%%%%%%%

As indicated above, the newborn NS formed in the core-collapse process will rapidly rotate, with their rotation rate limited by the mass-shedding angular velocity. We now estimate the NS angular velocity at the system's estimated age. For this task, we consider the NS can spin down as it might be subjected to the torque by magnetic dipole radiation. In this case, the energy conservation equation reads
\begin{equation}
	-\dot{E}_{\rm dip}\approx -\frac{d}{dt}\left(\frac{1}{2}I \Omega^2\right)=L_{\rm dip} = \frac{2}{3}\frac{B^2 R_{\rm NS}^6 \Omega^4}{c^3},
	\label{eq:dipole}
\end{equation}
where $I$ is the NS moment of inertia. Neglecting the change with time of $I$, Eq. (\ref{eq:dipole}) leads to the angular rotation evolution 
\begin{equation}
	\Omega(t)=\frac{\Omega_0}{\sqrt{1+\frac{t}{\tau}}},\qquad \tau=\frac{3 Ic^3}{4 B^2 R_{\rm NS}^6 \Omega_0^2},\label{eq:Omega}
\end{equation}
where $\tau$ is the spin-down timescale and $\Omega_0$ is the initial NS angular velocity, which we adopt to be a maximum value, i.e., $\Omega_0=\Omega_K \approx 0.7\sqrt{G M_{\rm NS}/R_{\rm NS}^3} \approx 7110$ rad s$^{-1}$, corresponding to a rotation frequency $f_0 = \Omega_0/(2\pi) \approx 1131$ Hz ($\approx 0.9$ ms period).
This is just the initial frequency at the birth of the NS, and it will evolve rapidly.

CCOs are characterized by weak magnetic fields $B\sim 10^{10}$--$10^{11}\rm G$. For instance, assuming $B = 10^{11}$ G, $M_{\rm NS}=0.9M_\odot$, $R_{\rm NS} = 10^6$ cm, and $I=10^{45}$ g cm$^2$, Eq. (\ref{eq:Omega}) tells that at $t = 4.5$ kyr, the NS rotation frequency will be $\approx 531$ Hz. This is an upper limit for the rotation frequency since additional effects could also contribute to the NS spin-down, e.g., the gravitational waves in early evolution \citep[see, e.g.,][]{1969ApJ...157.1395O,1969ApJ...158L..71F, RWlincei1970,1970ApJ...161..571C,1974ApJ...187..609M}, multipolar magnetic field components \citep[see, e.g.,][]{2013Natur.500..312T,2013MNRAS.434.1658M,2015MNRAS.450..714P,2016MNRAS.456.4145R,2019LRCA....5....3P,2022ApJ...939...62R,2023ApJ...945...95W}, or the magnetic field could have been larger at earlier times at then be buried by fallback accretion \citep[see, e.g.,][]{2011MNRAS.414.2567H,2018PhRvD..98h3012F}. Thus, the rotation effect on the NS structure at these times is negligible, so a slow-rotation or non-rotation approximation may suffice to estimate the mass and radius \citep[see, e.g.,][]{2015PhRvD..92b3007C}. 
%
% \begin{figure}
% \centering\includegraphics[width=0.5\textwidth]{./figures/spin_down.png}
% \caption{Time evolution of the NS rotational angular velocity assuming dipole radiation. Parameters are set to $M_{\rm NS}=0.9M_\odot$, $R_{\rm NS}= 10^6\rm cm$, and $I= 10^{45}\ \rm g cm^2$. The two vertical lines are $4.5$ and $10$ kyr, respectively.} 
% \label{fig:Spindown}
% \end{figure}

% Combining the information from the shell size, the post-AGB phase of the central star, and the X-ray bright phase of the SNR, it is inferred that IRAS 17287--3443 and XMMU J1732 were likely formed a binary disrupted by the SN explosion $\sim 4$--$10$ kyr ago \citep{2016MNRAS.458.2565D}. Our obtained deceleration timescale further strengthens this hypothesis, indicating that the NS has undergone at least several decay timescales (several kyr), resulting in a significant reduction in rotation rate, and it is now a slowly rotating NS.

We can also use the above estimate to constrain the magnetic field, the frequency, and the X-ray pulsar efficiency, as follows. As for other CCOs, the X-ray observations of XMMU J1732 by XMM-Newton show a stable, i.e., absent of pulsed, emission. The observed flux in the $0.5$--$10$ keV energy band is $F_X \approx 2.5 \times 10^{-12}$ erg s$^{-1}$ cm$^{-2}$ \citep{2015A&A...573A..53K}. The spectrum is consistent with a blackbody emission with surface temperature $T_s=2\times 10^6$ K and radius $R_{\rm NS} \approx 10.5$ km, assuming a distance to the source $d =2.5$ kpc \citep{2021AJ....161..147B}. This implies an intrinsic X-ray luminosity $L_X = 4\pi d^2 F_X \approx 2\times 10^{33}$ erg s$^{-1}$. The dipole luminosity in the X-ray band is $L^{\rm dip}_X = \eta_X L_{\rm dip}$, where $\eta_X$ is the X-ray emission efficiency parameter. Therefore, the constraint $L^{\rm dip}_X\leq L_X$ leads to
\begin{equation}
    \eta_X \leq \frac{3}{2} \frac{c^3 L_X}{B^2 R_{\rm NS}^6 \Omega_0^4}\left(1+\frac{t}{\tau}\right)^2,
\end{equation}
which for the above parameters implies $\eta_X \lesssim 10^{-7}$ at $t=4.5$ kyr. This value is consistent with the lowest X-ray emission efficiencies of pulsars  \citep[see, e.g.,][]{2013ApJS..208...17A}. Still, the value of $\eta_X$ can be higher since, as we argued above, the rotation angular velocity is an upper limit, so the CCO is likely slower.

%%%%%%%%%%%%%%%%%%%%%%%%%%%%%%%%%%%%%%%%%%%%%%%%%%%%%%
\subsection{Cooling evolution}\label{sec:2.4}
%%%%%%%%%%%%%%%%%%%%%%%%%%%%%%%%%%%%%%%%%%%%%%%%%%%%%%

We must also dedicate some time to exploring the thermal properties of XMMU J1732. This object offers a unique opportunity for examining its thermal characteristics, as it is one of the few thermally emitting compact objects for which we have a reliable mass estimate. 

We start by revisiting the thermal evolution equations that govern the cooling process of NSs. This cooling is primarily driven by the emission of neutrinos, originating from the star's interior, and photons, emitted from its surface. The equations representing both energy balance and conservation are written as
\begin{eqnarray}
  \frac{ \partial (l e^{2\phi})}{\partial m}& = 
  &-\frac{1}{\rho \sqrt{1 - 2m/r}} \left( \epsilon_\nu 
    e^{2\phi} + c_v \frac{\partial (T e^\phi) }{\partial t} \right) \, , 
  \label{coeq1}  \\
  \frac{\partial (T e^\phi)}{\partial m} &=& - 
  \frac{(l e^{\phi})}{16 \pi^2 r^4 \kappa \rho \sqrt{1 - 2m/r}} 
  \label{coeq2} 
  \, ,
\end{eqnarray}
Equations (\ref{coeq1}) and (\ref{coeq2}) illustrate that the cooling of NSs depends on both macroscopic properties—such as radial distance ($r$), mass ($m(r)$), and the metric function ($\phi(r)$). Additionally, there is a direct correlation with microscopic and thermodynamic quantities, including specific heat $c_v(r,T)$, thermal conductivity $\kappa(r,T)$, neutrino emissivity $\epsilon_\nu(r,T)$, and energy density $\rho(r)$.
By solving equations (\ref{coeq1}) and (\ref{coeq2}), one can obtain the temporal evolution of the temperature ($T(r,t)$) and luminosity ($l(r,t)$), which can then be compared against observational data to gauge the quality of the underlying model.

The stellar microscopic composition is crucial for thermal evolution, significantly impacting thermal conductivity, specific heat, and most importantly, neutrino emissivity. Given the nature of our model, we have opted for a conservative approach by using a parametrization of the Akmal-Pandaripande-Ravenhall (APR) EOS \citep{Akmal1998,Heiselberg2000}, known for its \textit{ab initio} formulation. This choice is ideal for modeling low-mass stars, as it provides reliable results for low densities, thereby minimizing uncertainties associated with the microscopic model. Previous studies have employed similar methodologies, such as \citep{2023ApJ...958...49S,PhysRevC.107.025806,2020A&A...642A..42S}.
Assuming the APR EoS, the composition of the stellar core is limited to neutrons, protons, and electrons. The corresponding microscopic properties are utilized to determine all pertinent thermodynamic quantities. Additionally, we consider all potential neutrino emission processes, including the direct Urca, modified Urca, and Bremsstrahlung processes.

The crust of the star is modeled using the traditional Baym-Pethick-Sutherland (BPS) approach \citep{Baym:1971pw}. In this model, the outer crust consists of heavy ions arranged in a crystalline lattice permeated by electrons. The inner crust begins at the neutron drip density, where, in addition to the electron sea, free neutrons are also present. Thermodynamics of the crust is mostly dominated by the heavy ions and electrons, the latter being responsible for most of the specific heat, with the latter driving most of the heat conduction.

Details of the calculations of thermodynamics properties may be found in great details in references \cite{Yakovlev2000,Yakovlev2004,Page2004}.

Finally, we must enforce the appropriate boundary conditions. These include the vanishing heat flow at the star's center ($l(r=0) = 0$) and the suitable atmospheric model at the surface, which depends on the ratio and content of light and heavy elements. For more details, refer to \cite{Gudmundsson1982, Gudmundsson1983, Page2006}. 

\begin{figure}
\centering\includegraphics[width=0.5\textwidth]{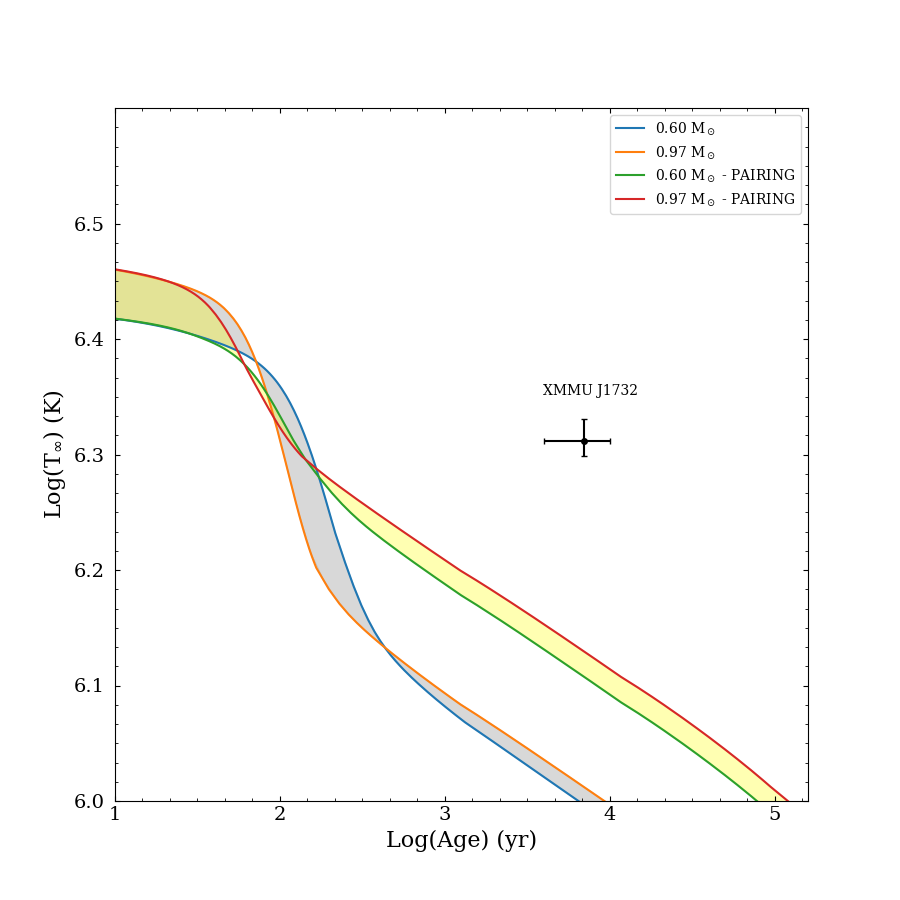}
\caption{Thermal evolution simulation for $0.6$--$0.97 M_\odot$ NSs modeled with the APR EOS. Cooling tracks with pairing indicated were calculated employing neutron singlet and triplet pairing ($^1S_0$ and $^3P_2$) as well as proton singlet ($^1S_0$) model after the SFB and CCDK models \citep{Schwenk:2002fq,Chen:1993bam}. These simulations considered an envelope of heavy elements. Also shown is the observed temperature of XMMU J1732 and its estimated age according to \citet{2016MNRAS.458.2565D}.} 
\label{fig:cool_HE}
\end{figure}

Our initial findings are presented in Fig.~\ref{fig:cool_HE}, which depicts the thermal evolution of neutron stars with masses in the range of $0.77_{-0.17}^{+0.2} M_{\odot}$ against the observed temperature of XMMU J1732 whose age was estimated in  \citep{2016MNRAS.458.2565D}, this age of between 4 –- 10 kyr is supported by the fact that the observed infrared structure and SNR shell must have comparable ages. Alternatively, we could adopt a more stringent age constraint of 2 to 6 kyr, as suggested by \cite{2016A&A...591A..68C}, which we also examined. However, this narrower age range did not lead to qualitative differences in our results, except that it slightly favors an object with a smaller mass, closer to 0.6 M$_\odot$. This figure includes two datasets: one for NSs without pairing and another with pairing.
Pairing is a crucial factor in the thermal evolution of neutron stars. Its significance to cooling lies in the exponential suppression of neutrino emission processes by the pairing gap energy, which affects the primary heat sink in the thermal evolution. At the densities present in neutron stars, nucleons are expected to form pairs \citep{Chen1993}. Although there are still many uncertainties regarding the strength and prevalence of pairing, it is generally accepted (see for instance Refs.~\cite{Page2004, Beloin:2016zop}) that, at least for lower densities, neutrons may pair up in singlet ($^1S_0$) states, especially in the lower density regions of the outer crust, as well as triplet ($^3P_2$) states, which can extend into the core. More uncertain, but still possible, is the formation of superconducting protons via singlet proton pairing.
We examine both neutron singlet and triplet pairing ($^1S_0$ and $^3P_2$), as well as proton singlet ($^1S_0$) based on the SFB and CCDK models \citep{Schwenk:2002fq,Chen:1993bam}. Allowing nucleons to pair up results in significantly slower cooling, yet it cannot match the observed temperature of XMMU J1732. As previously noted, the composition of a neutron star's atmosphere greatly influences its cooling process. Initially, our simulations assumed an atmosphere composed exclusively of heavy elements. Such a setup, which is associated with more efficient cooling, does not align with the specific conditions we propose for XMMU J1732 in this paper.

\begin{figure}
\centering\includegraphics[width=0.5\textwidth]{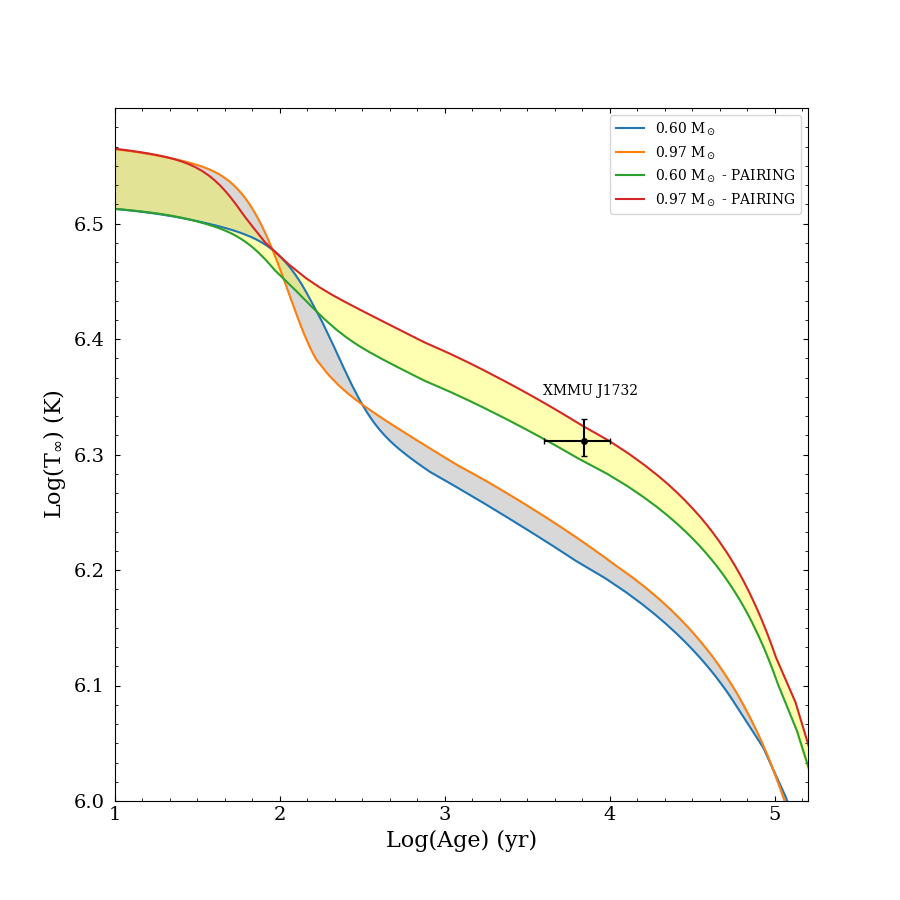}
\caption{Same as Fig.~\ref{fig:cool_HE} but for a hydrogen-rich envelope with a light-element fraction of  $\Delta M/M = 10^{-7}$, which is consistent with a carbon rich envelope.} 
\label{fig:cool_LE}
\end{figure}

The situation significantly improves when considering a light-element-rich envelope. In this case, we have employed an envelope with $\Delta M/M = 10^{-7}$, where $\Delta M$ represents the light elements mass in the upper envelope \citep{Gudmundsson1982,Gudmundsson1983,Potekhin1997}.  Figure~\ref{fig:cool_LE} shows the outcomes for such an envelope.

The selection of $\Delta M/M = 10^{-7}$ is deliberate, aligning with an atmosphere composed of H, He, and C as described by \cite{Potekhin1997}.  This detail is crucial since the most likely scenario for this object is the accretion of material onto the CCO shortly after the supernova, resulting in a significant carbon presence in the atmosphere, as it was pointed out by \citep{2016MNRAS.458.2565D}. It is thus worth noting that, by accounting for a carbon-rich atmosphere, there is a remarkable concordance between the temperature and estimated age of the central compact object XMMU J1732 and our cooling simulations.

The results suggest the thermal properties of XMMU J1732 concur with the model explored in this study. An ordinary NS with a mass ranging $0.6$--$0.97 M_\odot$ can naturally account for the observed data. In this study, we have focused solely on hadronic degrees of freedom. This approach is reasonable because transitions to quark matter are suggested to occur at higher densities \cite{Oertel2017,Alford2013,Buballa2014,Masuda2013} than those expected in a compact object with such low mass. Notably, the work published in \cite{2023ApJ...958...49S,horvath2023light} has demonstrated that exotic degrees of freedom can also explain the thermal data of HESS J1731--347. Therefore, extending our evolutionary model to include exotic degrees of freedom would be an interesting direction for future research.

%%%%%%%%%%%%%%%%%%%%%%%%%%%%%%%%%%%%%%%%%%%%%%%%%%%%%%%
%%%%%%%%%%%%%%%%%%%%%%%%%%%%%%%%%%%%%%%%%%%%%%%%%%%%%%%
\section{Progenitor system} \label{sec:3}
%%%%%%%%%%%%%%%%%%%%%%%%%%%%%%%%%%%%%%%%%%%%%%%%%%%%%%%
%%%%%%%%%%%%%%%%%%%%%%%%%%%%%%%%%%%%%%%%%%%%%%%%%%%%%%%

Having clarified the consistency of XMMU J1732 with a rotating core-collapse event that shed mass, we turn to reconstruct the astrophysical scenario before the SN event. Observations indicate that the mass of the dust shell in this SNR is $\sim 1.5$--$3M_\odot$, which further supports the suggestion that the pre-SN progenitor should not be a single star but rather an ultra-stripped core in a binary. Below, we infer the progenitor binary parameters.

%%%%%%%%%%%%%%%%%%%%%%%%%%%%%%%%%%%%%%%%%%%%%%%%%%%%%%
\subsection{Binary System Prior to Supernova}\label{sec:3.1}
%%%%%%%%%%%%%%%%%%%%%%%%%%%%%%%%%%%%%%%%%%%%%%%%%%%%%%

\begin{figure}
\centering\includegraphics[width=0.5\textwidth]{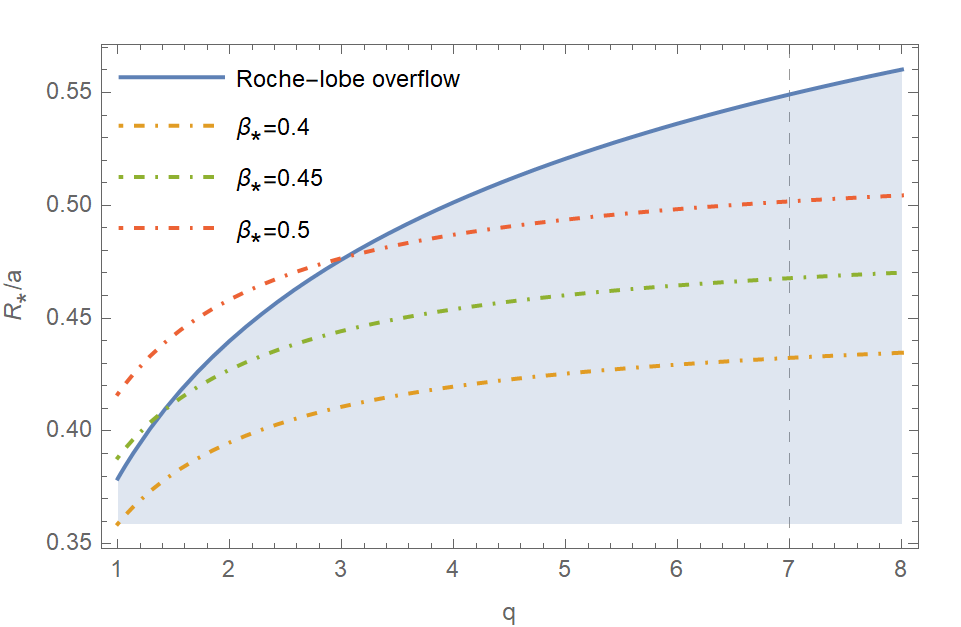}
\caption{The ratio of the primary star radius to the binary semi-major axis as a function of the ratio of the masses of the two components of the binary star. The blue curve corresponds to the Roche-lobe outflow condition, with the shaded region separating two components larger than the Roche-lobe outflow condition, i.e., Eq. (\ref{eq:Roche}). The dashed-dot curves correspond to the tidally locked condition with different rotation parameters. The vertical dashed line indicates the mass ratio of the inferred progenitor binary associated with SNR HESS J1731--347.} 
\label{fig:BinaryConstrant}
\end{figure}

We start by imposing that there is no Roche-lobe overflow in the binary at the SN event. For the primary star, this condition constraints its radius $R_\star$ to satisfy
\begin{equation}
	\frac{R_\star}{a} \lesssim \frac{0.49 q^{2/3}}{0.6q^{2/3}+\ln(1+q^{1/3})},
	\label{eq:Roche}
\end{equation}
where  $a$ is the semi-major axis of the binary orbit, and $q \equiv M_\star/M_{\rm c} > 1$ is the binary mass ratio. When the equal sign in Eq. (\ref{eq:Roche}) is taken, it indicates the Roche-lobe outflow condition, corresponding to the blue curve in the Fig. \ref{fig:BinaryConstrant}, with the shaded region indicating the range defined by the Eq. (\ref{eq:Roche}).

We assume the binary is tidally locked before the SN occurs, i.e., $\Omega_\star = \Omega_{\rm orb}$. Then, the angular velocity parameter is given by 
\begin{equation}
	\beta_\star \equiv \frac{\Omega_\star}{\Omega_{K,*}}= \frac{\sqrt{\frac{G(M_\star+M_{\rm c})}{a^3}}}{0.76 \sqrt{\frac{GM_\star}{R_\star^3}}}=\frac{1}{0.76}\sqrt{\frac{1+q}{q}}\left(\frac{R_\star}{a}\right)^{3/2}.
	\label{eq:Tidal}
\end{equation}
Note that the value $0.76$ is consistent with Section \ref{sec:2.2}. For a given $\beta_*$, Eq. (\ref{eq:Tidal}) relates $R_\star/a$ and $q$. As discussed in section \ref{sec:2.2}, assuming that the collapsing iron core rotates uniformly to collapse into a low-mass NS, a reasonable value for $\beta_\star$ could be somewhere in the range $0.4$--$0.5$. Figure \ref{fig:BinaryConstrant} shows the $R_\star/a$ as a function of $q$ for three selected values, $\beta_* = 0.4$, $0.45$, and $0.5$.

An SNR mass of $3M_\odot$ \citep{2016MNRAS.458.2565D} implies the pre-SN star would be $M_* = 4.2 M_\odot$. We recall that here we refer to the pre-SN star mass, which can be considerably lower than the mass of its progenitor zero-age main-sequence star. Indeed, for the latter to give rise to an NS, stellar evolution predicts it to be $\gtrsim 8 M_\odot$. Therefore, the above numbers suggest the star loses about half of its mass from the main sequence to the pre-SN stage. Given that the companion star is $M_c \approx 0.6 M_\odot$, the mass ratio is $q\approx 7$. These parameters are consistent with the ultra-stripped SNe simulations: see, for instance, the simulation where the mass of the stripped star is $4.23 M_\odot$ and the iron core mass is $1.27 M_\odot$ in Table 1 in \citet{2021A&A...656A..58L}. The stripped star of $4.23 M_\odot$ has a stellar radius $R_\star\approx 4.2 R_\odot$. By substituting this value into Eq. (\ref{eq:Tidal}), for $q=7$, we find the orbital separation $a=9.04 R_\odot$. This implies an orbital period of $P=1.43$ days. The optical star has $M_c\sim 0.6 M_\odot$, so its radius is expected to be $\lesssim R_\odot$ during the post-AGB phase. Thus, at the time of the SN explosion, the separation between the two components is larger than the sum of their radii, placing it below the Roche-lobe outflow condition curve, as shown in Fig. \ref{fig:BinaryConstrant}. Interestingly, this also agrees with the numerical simulations of \citet{2021A&A...656A..58L}, which indicate that the primary star in close binaries at solar metallicity and an initial mass above $11M_\odot$ does not interact with its companion after reaching core helium depletion.

%%%%%%%%%%%%%%%%%%%%%%%%%%%%%%%%%%%%%%%%%%%%%%%%%%%%%%%
\subsection{Binary Disruption} \label{sec:3.2}
%%%%%%%%%%%%%%%%%%%%%%%%%%%%%%%%%%%%%%%%%%%%%%%%%%%%%%%

We now show that the binary parameters inferred above imply that the SN explosion likely disrupted the binary, as suggested by \citet{2016MNRAS.458.2565D}. The SNR mass consists of two parts: the ejecta from the SN explosion, $M_{\rm ej}=4.23 M_\odot$--$1.27 M_\odot=2.96 M_\odot$ (the outer layers above the iron core), and the material shed during the core collapse, $M_{\rm shed}\approx 0.3M_\odot$ (i.e., $M_{\rm cco}+M_{\rm shed}\approx 1.27 M_\odot$). It is worth noting that the mass ejected from the AGB to post-AGB evolution of the optical star is expected to be only $10^{-5}$--$10^{-2.5} M_\odot$ \citep{2003A&A...405.1075M,2012MNRAS.424.2345V}, hence we neglect its contribution to the remnant mass.

For a binary with a circular orbit, an ejected mass larger than half of the total mass, i.e., $M_{\rm ej}>M_{\rm tot}/2$, leads to a $100\%$ of probability of disruption \citep{1983ApJ...267..322H}. As indicated by the model above, $M_{\rm ej}=2.96 M_\odot$ and $M_{\rm tot}=4.23 M_\odot+0.6M_\odot=4.83 M_\odot$. Therefore, $M_{\rm ej}/M_{\rm tot}>1/2$, indicating that the SN explosion could have indeed disrupted this binary. Furthermore, the mass of the SNR is consistent with the upper limit of approximately $3 M_\odot$ \citep{2016MNRAS.458.2565D}.

Following \citet{1983ApJ...267..322H}, we can estimate an upper limit of the kick velocity, i.e., assuming $100\%$ of the probability of disruption. This leads to a kick of $\Delta v\approx 670$ km s$^{-1}$. The possible large kick velocity imparted to the stars in this scenario agrees with existing simulations of $\sim 1$ day orbital period binaries of low-mass components \citep{2004ApJ...612.1044P}. Considering the SNR estimated age of $4.5$ kyr, the two objects could have reached a maximum separation of $\approx 3$ pc. This value is larger than the observed relative projected offset of $\approx 0.3$ pc between the CCO, XMMU J1732, and the optical star, IRAS 17287--3443, which is reasonable because we have used the upper limit of the kick velocity, and the actual distance between the optical star and the CCO can be larger than the observed projected offset, depending on the proper motion direction relative to the line of sight.

%%%%%%%%%%%%%%%%%%%%%%%%%%%%%%%%%%%%%%%%%%%%%%%%%%%%%%%%%%%
%%%%%%%%%%%%%%%%%%%%%%%%%%%%%%%%%%%%%%%%%%%%%%%%%%%%%%%%%%%
\section{Discussion and Conclusions} \label{sec:4}
%%%%%%%%%%%%%%%%%%%%%%%%%%%%%%%%%%%%%%%%%%%%%%%%%%%%%%%%%%%
%%%%%%%%%%%%%%%%%%%%%%%%%%%%%%%%%%%%%%%%%%%%%%%%%%%%%%%%%%%

We have advanced a formation scenario for the puzzling, light CCO XMMU J1732, with an optical star neighbor IRAS 17287--3443 within the SNR HESS J1731--347. We have provided a scenario within the traditional framework of NS formation from core-collapse SN. Here are some concluding remarks:

(i) Angular momentum conservation in the collapse process leads to the light NS formation by mass-shedding if the iron core at the collapse moment rotates at about $45\%$ of the Keplerian limit angular velocity, i.e., $\beta \sim 0.45$. The observed CCO, XMMU J1732, is indeed a light NS of $M_{\rm NS}\approx 0.9 M_\odot$ formed in the collapse of a fast rotating iron core of mass $M\approx 1.2 M_\odot$, with a rotation period of $\approx 7$ s, which sheds $\approx 0.3 M_\odot$ during its collapse avoiding to overcome the maximum angular momentum the newborn NS can hold. We refer to section \ref{sec:2.2} for details.

(ii) Assuming a magnetic dipole braking model for a dipole strength of $10^{11}$ G, and as NS age the estimated age of the SNR ($4.5$--$10$ kyr), we showed the CCO must be currently a modest rotator, given the upper limit of the rotation frequency $\approx 531 $ Hz (see section \ref{sec:2.3}). Thus, a slow-rotation or non-rotating mass-radius relation could accurately describe it. This implies that the observationally inferred radius of XMMU J1732 can be a relevant constraint in the low-mass region of the NS mass-radius relation.

%\textcolor{red}{(iii) HERE CONCLUSION ON COOLING CALCULATION.}
(iii) We have performed comprehensive cooling simulations for light neutron stars within the mass range of \(0.77_{-0.17}^{+0.20} M_{\odot}\), specifically for those characterized by the \textit{ab-initio} APR model, and have observed a significant agreement with empirical data. It is important to highlight that our simulations incorporated considerations for nucleon pairing, which align with the prevailing theories on NS thermal evolution. Remarkably by considering a carbon-rich atmosphere, as predicted by \citep{2016MNRAS.458.2565D}, our cooling simulations were in excelent agreement with observed thermal data. This concurrence is not merely coincidental but is a testament to the robustness of the the underlying assumptions. These findings corroborate the model introduced in this study, demonstrating that the thermal characteristics of XMMU J1732 can be accurately accounted for by a conventional NS cooling model, thereby eliminating the need for more speculative hypotheses and decreasing the number of parameters required to interpret the observed phenomena.

(iv) The obtained parameters from the core collapse model are generally consistent with simulations of the pre-SN stage in binaries with the primary having a mass of $M_*\sim 4.2 M_\odot$ \citep{2021A&A...656A..58L}. We assume XMMU J1732 and IRAS 17287--3443 ($M_c \approx 0.6 M_\odot$) formed a binary system before the core-collapse SN event. Adopting tidal locking, we have inferred a semi-major axis of the pre-SN binary $a\approx 9 R_\odot$, and the orbital period $P\approx 1.4$ days (see section \ref{sec:3.1}). No Roche-lobe overflow occurred before the SN.

(v) The mass loss amounting to $\sim 3 M_\odot$, given by the SNR shell mass, led to the disruption of the binary since it is more than half the total binary mass, $M_* + M_c \approx 4.8 M_\odot$. This is consistent with the observed projected separation of $\approx 0.3$ pc between the CCO and the post-AGB, lower than the maximum separation of $\approx 3$ pc, obtained assuming the maximum possible kick ($670$ km s$^{-1}$) that could have been imparted (see section \ref{sec:3.2}).

Although assessed via a simplified model, the above astrophysical scenario highlights the relevance of rotation in the binary evolution and, finally, in the core-collapse process. Accurate calculations from numerical simulations could replace some simplifications. For instance, we have made a \textit{hybrid} model of the rotating core collapse that joins a simplified Newtonian model to the results of the rotating NS structure obtained in full general relativity. Thus, it would be ideal to perform a full general relativistic calculation of the gravitational collapse of a rotating iron core of a pre-SN star with its interior structure obtained with an evolution code accounting for binary interactions.

To conclude, some comments on the binary evolution path are in order. Typically, ultra-stripped SNe do not lead to the binary disruption \citep{2015MNRAS.451.2123T,2023Natur.614...45R}. However, what makes this system unique is mainly the exceptionally low mass of the newborn NS, i.e., the CCO, and the companion star, resulting in the ejected mass easily exceeding half of the total system mass. The previous evolution of the binary, especially of the low-mass companion, remains an interesting subject of study since the evolution leading to a post-AGB is expected to be longer than that leading to a core-collapse SN and an NS. We can only speculate that our binary could have followed a similar evolutionary path to the \citet{2021A&A...656A..58L} simulations. The crucial condition is that the binary mass ratio be close to unity at the beginning of the evolution. Their simulations involve a secondary with an initial mass of $80\%$ of the primary star. Still, the secondary's state at the time of the SN is uncertain. Therefore, further population synthesis analyses and simulations of the binary stellar evolution, including possible different masses of the secondary, are needed to comprehend this system fully. Such analysis, which is worth it on its own, goes beyond our scope here and is left for future work. 

\newpage
\begin{acknowledgments}
We thank Prof. Thomas Tauris for discussing and highlighting important references to binary evolution and Prof. M. Bulla for suggestions on the presentation. J.A.R. also thanks Prof. M. Malheiro and Prof. A. Drago for discussions on the quark star hypothesis of XMMU J1732. 
SRZ acknowledges the Joint International Relativistic Astrophysics Doctorate Program and support from the China Scholarship Council (CSC No. 202206340085). R.N. acknowledges financial support from CAPES and FAPERJ. This work is part of the project FAPERJ JCNE Proc. No. E-26/201.432/2021. 
\end{acknowledgments}

% \appendix

% %%%%%%%%%%%%%%%%%%%%%%%%%%%%%%%%%%%%%%%%%%%%%%%%%%%%%%%%%%%%%%%
% %%%%%%%%%%%%%%%%%%%%%%%%%%%%%%%%%%%%%%%%%%%%%%%%%%%%%%%%%%%%%%%
% \section{APPENDIX NAME} \label{sec:app}
% %%%%%%%%%%%%%%%%%%%%%%%%%%%%%%%%%%%%%%%%%%%%%%%%%%%%%%%%%%%%%%%
% %%%%%%%%%%%%%%%%%%%%%%%%%%%%%%%%%%%%%%%%%%%%%%%%%%%%%%%%%%%%%%%
% We start 

\bibliography{sample}{}
\bibliographystyle{aasjournal}

\end{CJK*}
\end{document}